\documentclass[amssymb,aps,prl,twocolumn]{revtex4}
\usepackage{amssymb}
\usepackage{graphicx}
\usepackage{dcolumn}
\usepackage{bm}
\usepackage{amsmath}
\begin{document}

\title{High sensitivity phonon spectroscopy of Bose-Einstein
  condensates \\ using matter-wave interference}

\author{N. Katz$^1$, R. Ozeri$^1$\footnote{Current address: Time and Frequency Division
NIST 325 Broadway Boulder, Colorado 80305, USA}, J.
Steinhauer$^1$\footnote{Current address: Department of Physics,
Technion -- Israel Institute of Technology, Technion City, Haifa
32000, Israel} , N. Davidson$^1$, C. Tozzo$^{2,3}$ and F.
Dalfovo$^{3,4}$ }

\affiliation{$^1$ Department of Physics of Complex Systems,
Weizmann Institute of Science, Rehovot 76100, Israel \\
$^2$ Dipartimento di Fisica, Universit\`a di Trento, I-38050 Povo, Italy \\
$^3$ Istituto Nazionale per la Fisica della Materia, BEC-INFM Trento,
I-38050 Povo, Italy \\
$^4$ Dipartimento di Matematica e Fisica, Universit\`a Cattolica
del Sacro Cuore, Via Musei 41, 25121 Brescia }

\begin{abstract}

We study low momentum excitations of a Bose-Einstein condensate
using a novel matter-wave interference technique. In time-of-flight
expansion images we observe strong matter-wave fringe patterns. The
fringe contrast is a sensitive spectroscopic probe of in-trap phonons
and is explained by use of a Bogoliubov excitation projection method
applied to the rescaled order parameter of the expanding condensate.
Gross-Pitaevskii simulations agree with the experimental data and
confirm the validity of the theoretical interpretation. We show that
the high sensitivity of this detection scheme gives access to the
quantized quasiparticle regime.

\end{abstract}

\pacs{03.75.Kk, 03.75.Lm, 32.80.-t}

\maketitle

The peculiar nature of a Bose-Einstein condensate (BEC) is clearly
observed in its low momentum excitations, revealing the many-body
coherent nature of the system. Recent experiments have provided
observations of the Bogoliubov excitation spectrum, superfluidity,
and suppression of low momentum excitations due to quantum
correlations in the ground state
\cite{Inhomogeneous,excitation_spectrum,enhancement}.

In these experiments the condensates were typically excited by
Bragg pulses \cite{Ketterle_Doppler} and imaged after a free
expansion. At high momenta the excitations were clearly separated
from the expanding condensate, and could be counted and
quantified. Even in the phonon regime (wavelength larger than but
comparable to the healing length) a well defined excitation cloud
could still be distinguished from the condensate (see Fig.~1a) and
was found to be amenable to direct atom counting methods.

However, at sufficiently low momentum (wavelength much larger than
the healing length), we observe a new regime, in which the
excitations and the condensate no longer separate, regardless of
the duration of the expansion. The excitations are manifested in a
clear density modulation of the cloud in the absorption images
(see Fig.~1c), reminiscent of in-situ observations of low energy,
zero momentum, excitations of the condensate
\cite{parametric_ketterle,parametric_wieman}, but they differ
significantly from these since they involve both nontrivial
time-of flight (TOF) dynamics and a small, but finite, excitation
momentum.

\begin{figure}[h]
\begin{center}
\includegraphics[width=8cm]{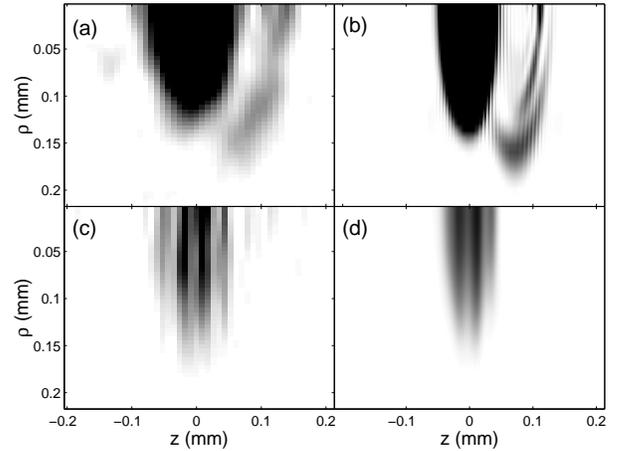}
\end{center}
\caption{Density $n(\rho,z)$ of the expanding condensate,
initially excited with phonons of momentum $q$. In (a) and (b):
$qa_{\rho}=2.03$. The density in (a) is obtained by computerized
tomography of TOF column density absorption images, $n_{\rm
col}(y,z)$, while (b) corresponds to a full GP simulation. In both
cases, note the clearly separated excitation cloud on the right of
the condensate. The same quantities are plotted in (c) and (d) for
$qa_{\rho}=0.31$. Note the strong density modulation, with no
significant outcoupled fraction.  } \label{fig1}
\end{figure}

In this Letter we analyze the dynamics of phonons in such freely
expanding condensates using both the Gross-Pitaevskii equation (GPE)
and a dynamically rescaled Bogoliubov theory \cite{theory_fringes}.
We find that, in our elongated condensate, axial low-momentum
phonons are adiabatically converted by the (mainly radial) expansion
into free atoms with the same axial momentum. The overlap of these
moving free particles with the expanding ground state results in
the axial periodic density modulations observed in the TOF images.

We use the fringe visibility of these density modulations after
TOF, as an extremely sensitive spectroscopic probe of the
excitation strength. Whereas the sensitivity of atom or momentum
counting methods \cite{excitation_spectrum} scales quadratically
with the quasiparticle excitation amplitude, we show that the
fringe visibility scales linearly with these amplitudes. This
enhanced sensitivity was noted and used to estimate the degree of
adiabaticity in the loading of a BEC into an optical lattice
\cite{phillips_lattice}, but not as a spectroscopic probe of the
excitations themselves.

Our nearly pure ($>95\%$) $^{87}$Rb condensate in the
$|F,m_{f}\rangle=|2,2\rangle$ ground state, is formed in an
elongated cylindrically-symmetric harmonic trap with axial
frequency $\omega_z=2\pi\times 25$ Hz and radial frequency
$\omega_{\rho}=2\pi\times 220$ Hz. The radial harmonic oscillator
length is $a_{\rho}=(\hbar/m\omega_{\rho})^{1/2}= 0.73 \mu$m. We
use condensates with $N=10^5$ atoms, having chemical potential
$\mu=\eta\hbar\omega_{\rho}$, with $\eta= 9.1$, and an average
healing length $\xi \sim 0.3 a_{\rho}$. Using Bragg excitation
beams detuned $6.5$ GHz above the $5S_{1/2},F=2\longrightarrow
5P_{3/2},F^{\prime }=3$ transition, Bogoliubov quasiparticles are
excited in the condensate along the $\hat{z}$ axis, at a
controlled wavenumber $q$ by varying the angle between the two
beams. At a given $q$ the frequency difference between the beams
$\omega$ is controlled via acousto-optic modulators. After the
Bragg pulse (of duration $t_B=6$ $m$sec) the magnetic trap is
rapidly turned off and after $38$ $m$sec of TOF an on-resonant
absorption image is taken, with the absorption beam perpendicular
to the $\hat{z}$-axis.

The dynamics of phonons within an expanding condensate has
recently been studied ~\cite{theory_fringes} and shown that,
within a short time interval ($t \sim \omega_\rho^{-1}$), phonons
are converted into single particle excitations travelling at an
axial velocity of the order of $\hbar q/m$. This velocity should
be compared to that of the slowly expanding axial boundary, which
travels at the asymptotic velocity
$\pi\omega^2_zZ_{TF}/2\omega_{\rho}$, where $Z_{TF}= a_\rho
(\omega_\rho/\omega_z) (2\eta)^{1/2}$ is the Thomas-Fermi axial
radius of the condensate. Thus at long expansion times the
excitations can separate out from the condensate only if
$qa_{\rho}>q_{c}a_{\rho} = \pi (\omega_z /\omega_{\rho})(\eta
/2)^{1/2}$, while for $q<q_c$  they remain within the condensate
at all times. For our experimental parameters one has
$q_{c}a_{\rho}=0.76$. Experimentally, this estimate is remarkably
exact. Indeed, we begin to observe a distinct excitation cloud at
$qa_{\rho}\simeq 0.72$. However, this transition is not sharp and
a small fraction still overcomes the condensate boundary even at
lower $q$. This escaped fraction was used to find the resonance in
our previous measurement of the excitation spectrum
\cite{excitation_spectrum}.

\begin{figure}[h]
\begin{center}
\includegraphics[width=8cm]{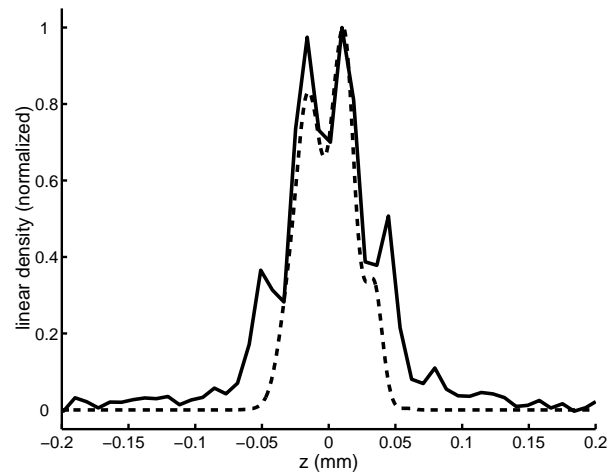}
\end{center}
\caption{Linear density $2\pi\int_{0}^{\infty}\rho
n({\rho},z)d\rho$ measured from TOF image with very low momentum
excitation $qa_{\rho}=0.31$ (solid line). Note the slight
asymmetry to right, which is in agreement with the GPE simulation
(dashed line).} \label{fig2}
\end{figure}

In the phonon regime, but with $q>q_{c}$, using computerized
tomography \cite{tomography} (see Fig.~1a), the freely expanding
excitation cloud was measured to be carrying almost all the
momentum and excess energy of the system due to excitation. By
solving the GPE for the TOF dynamics, we now confirm this
experimental observation quantitatively for this regime. For
instance, the lateral cloud in Fig. 1b carries about $99\%$ of the
momentum and more than $95\%$ of the energy of the initial
phonons. In Fig.~1b one also notices that the released cloud is
radially wider than the condensate, in agreement with the
experimental observation \cite{tomography} and has an evident
``shell structure". Both features are associated with the spatial
structure of the Bogoliubov amplitudes of the in-trap initial
excitations \cite{theory_fringes} and the presence of excitations
with radial nodal lines. At this momentum ($qa_{\rho}=2.03$)
counting the population of the excitation cloud also gives
reasonable agreement with the population expected by simple
structure factor considerations \cite{excitation_spectrum}.

Turning our attention to the very low momentum excitations with
$qa_{\rho}=0.31$, we study the TOF density modulations of Figs.~1c
and 1d in detail. Fig.~2 shows the linear density of the expanded
condensate, defined as $2\pi\int_{0}^{\infty}\rho n(\rho,z)d\rho$.
We note a slight asymmetry to the right, which is related to the
initial direction of propagation of the phonon\cite{comment1}.

\begin{figure}[h]
\begin{center}
\includegraphics[width=8cm]{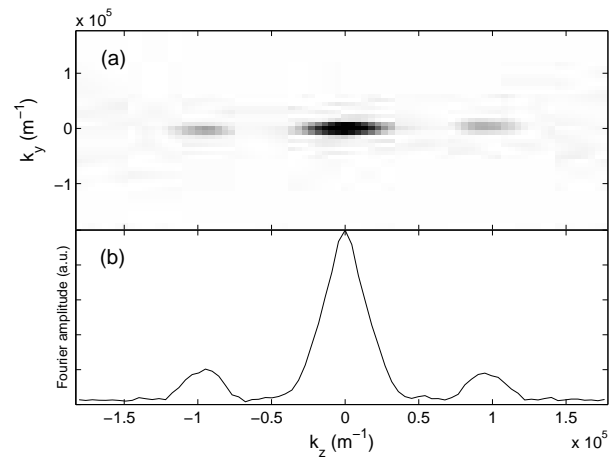}
\end{center}
\caption{Fourier transform of the column density $n_{\rm
col}(y,z)$ for the same image as Fig.~1c. (a) Fourier transform.
(b) Integration over central strip of (a). Note the clear
separation between the central mode and the sidelobes, despite the
low momentum of the excitation.} \label{fig3}
\end{figure}

To quantify the fringes of Fig.~1c, we take the Fourier transform
of the measured density profile, and observe clear sidelobes
(Fig.~3). We define the fringe visibility as the area of the
sidelobes divided by the area of the central lobe. This is a
powerful "spatial lock-in" method to reduce noise in the
measurement. It also agrees strictly with the standard definition
of contrast. We note the ease of the measurement, due to the clear
separation between the Fourier peaks, despite the very low momenta
involved. This should be contrasted with the atom counting
methods, where it becomes increasingly difficult to discern the
excitation cloud at finite expansion times in real space. At such
low momentum we do not expect any significant radial excitations,
since the coupling to all higher radial modes is predicted to be
negligible \cite{radial_modes,radial_theory}. This is confirmed by
the lack of any features in the transverse direction in Figs.~1c
and 3a.

Fig.~4 shows the fringe visibility as a function of the Bragg
frequency detuning (circles). We find the resonances to be at
$\pm137\pm10$ Hz, which compares favorably with the expected value
$138\pm5$ Hz for the average Bogoliubov frequency in local density
approximation (LDA) \cite{LDA}. The width of the resonances is due
to the finite duration of the Bragg pulse.

A direct GPE simulation of the experiment (Fig. 4, solid line)
gives good agreement with the observed data \cite{offset}.
However, little intuition is gained by this. Therefore, we also
compare this spectrum to the analytic result obtained with a
rescaled quasiparticle projection method \cite{theory_fringes}
(dashed line).

For this, we consider a uniform gas in a box of volume $L^3$, with
a density $n_0= N/L^3$. The ground state order parameter is
$\Psi_0= \sqrt{n_0}$, with chemical potential $\mu=gn_0$. The
excited states are the well known Bogoliubov quasiparticles,
having frequency $\omega_k = \{ k^2/(2m) [ \hbar^2 k^2/(2m) + 2
\mu]\}^{1/2}$ and amplitudes $u_k, v_k = \pm \{ [ \hbar^2 k^2/(2m)
+ 2 \mu]/(2\hbar \omega_k) \pm 1/2 \}^{1/2}$.

We insert the linear expansion
\begin{eqnarray}
\Psi(z,t) &=& e^{-i\mu t/\hbar}  \left\{  n_0^{1/2} + \right.
L^{-3/2} \sum_{k} c_k (t) u_k e^{i (kz - \omega_k t)}
\nonumber \\
&& + \left. c_k^*(t) v_k^* e^{-i(kz- \omega_k t) } \right\}
\label{eq:ansatz}
\end{eqnarray}
into the GPE. The Bragg pulse is included through the potential $V
= V_B \cos (qz - \omega t)$, acting for $-t_B < t < 0$. At
$t=-t_B$ the condensate is assumed to be in the ground state, so
that $c_k (-t_B)=0$ for all $k$'s. The equations for the
coefficients $c_k (t)$ are then solved analytically
\cite{radial_theory,blakie}. After the Bragg pulse one gets
\begin{equation}
c_{\pm q} (0) = \mp \frac{V_B \sqrt{N}(u_q+v_q)}{2 \hbar (\omega
\mp \omega_q)} \left( e^{\pm i(\omega \mp \omega_q)t_B}-1 \right)
\; , \label{eq:cpm}
\end{equation}
and $c_{k} (0) =0$ for $|k| \neq q$.

Next we assume that the uniform gas undergoes a fictitious radial
expansion which mimics the actual expansion, by taking a uniform
time-dependent density that decreases in time as the average
density of elongated condensate. For this we choose the density
$n_0$ to be equal to the radial average of the density of a
cylindrical condensate, $n_0=(2/5)n(0)$, having the same central
density $n(0)$ of our trapped condensate, and we treat the
expansion by means of the scaling ansatz \cite{kagan,castin}.

In this case, the density decreases as $n_0/b^2(t)$, where
$b^2(t)=1+\omega_\rho^2 t^2$ is the radial scaling parameter. In
the presence of quasiparticles with wavevector $\pm q$, one can
write a rescaled order parameter $\tilde{\Psi}$ in the form

\begin{equation}
\tilde{\Psi}(z,t)  =   \tilde{\Psi}_0 +  L^{-3/2} \sum_{k=\pm q}
c_k (0) \tilde{u}_k (t) e^{i kz } +  c_k^*(0) \tilde{v}_k^*(t)
e^{-ikz }  \; . \label{eq:ansatz2}
\end{equation}
At $t=0$, when $\tilde{u}_k(0)= u_k$, $\tilde{v}_k(0)=v_k$ and
$\tilde{\Psi}_0=\sqrt{n_0}$, this expression coincides with
(\ref{eq:ansatz}). At $t>0$, $\tilde{\Psi}_0$ remains stationary
while the quantities $\tilde{u}$ and $\tilde{v}$ are
time-dependent amplitudes that obey rescaled Bogoliubov-like
equations \cite{theory_fringes}.

\begin{figure}[h]
\begin{center}
\includegraphics[width=8cm]{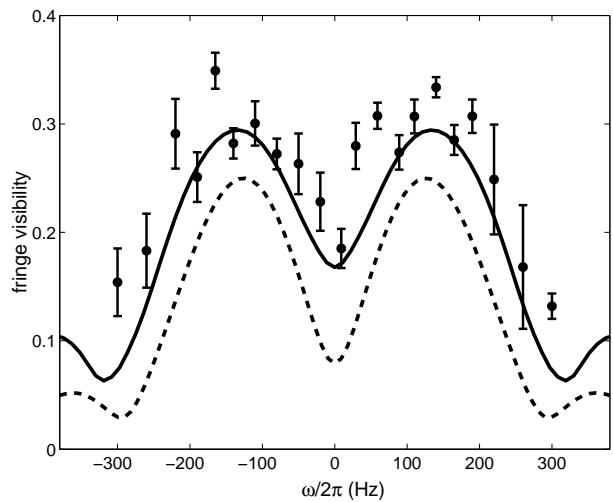}
\end{center}
\caption{Fringe visibility as a function of the Bragg excitation
frequency (points). The error bars represent the uncertainty due
to four measurements. We observe a clear double-peaked spectrum,
which is finite-time broadened. The peaks are found at $\omega=
\pm137\pm10$, close to the expected Bogoliubov frequency
($138\pm5$ Hz).The solid line is a full GPE simulation of the
experiment. The dashed line is the result of Eq.~(5).  }
\label{fig4}
\end{figure}

Further simplification is obtained by assuming adiabatic
quasiparticle-to-particle conversion. This corresponds to assuming
that, after an expansion time $t \gg \omega_\rho^{-1}$, the
rescaled amplitude $\tilde{u}$ becomes unity, while  $\tilde{v}$
vanishes \cite{note-adiabaticity}. In terms of $\tilde{\Psi}$,
this implies
\begin{equation}
\tilde{\Psi}(z,t) \to    \tilde{\Psi}_0  + L^{-3/2}  [ c_q (0)
e^{iqz} + c_{-q}^*(0) e^{-iqz } ] \; . \label{eq:asymptotic}
\end{equation}
Since the expansion is assumed to be radial, the linear density
obtained by radial integration of the rescaled density
$|\tilde{\Psi}|^2$ coincides with that obtained from
$|\Psi|^2$. The same is true for their Fourier components with
$k=0$ and $k= \pm q$. The sum of the $\pm q$ coefficients, divided
by the $k=0$ coefficient is exactly the visibility $\gamma$ as defined
for the experimental analysis. One finds $\gamma = (2/\sqrt{N})
|c_q(0)+ c_{-q}^*(0)|$ or, recalling
Eq.~(\ref{eq:cpm}),
\begin{eqnarray}
\gamma &=& \frac{V_B}{\hbar}\left[1+\frac{4\mu m}{\hbar^2q^2}
\right]^{-1/4} \nonumber\\
&\times&\left|\frac{e^{i(\omega-\omega_q)t_B}-1}{\omega_q-\omega}
+\frac{e^{-i(\omega+\omega_q)t_B}-1}{\omega+\omega_q}\right|\; .
\label{eq:adifrin}
\end{eqnarray}
The dashed line in Fig.~4 corresponds to this expression, calculated
for the same $V_B$ used in the GPE simulations ($V_B = 0.1 \hbar
\omega_\rho$) and with no fitting parameters. The agreement with
both experiment and GPE simulation indicates
that the essence of the release process of a BEC with such very
low momentum excitations is indeed captured by simply assuming
that, due to the reduction of the average background density in
which the excitations live, quasiparticles are adiabatically
converted into free particles that interfere with the ground
state.

\begin{figure}[h]
\begin{center}
\includegraphics[width=8cm]{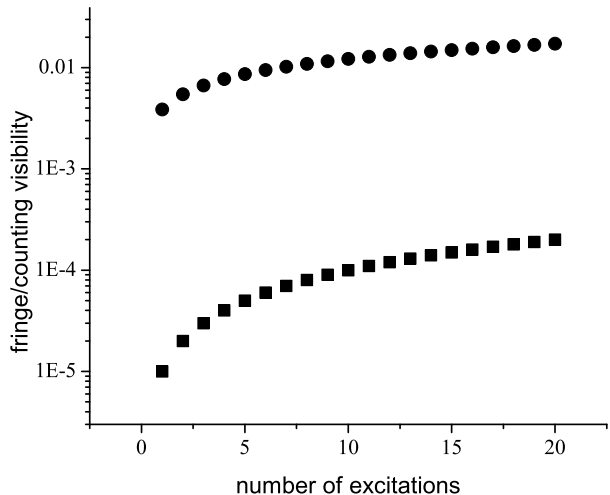}
\end{center}
\caption{Fringe visibility after TOF for $qa_{\rho}=0.31$
(circles) compared with  the standard atom counting technique for
$qa_{\rho}=2.03$ (boxes), as a function of the number of
excitations. The visibility of the atom counting method grows
linearly in excitation number with a very weak slope of $1/N$,
whereas the fringe visibility is not negligible ($\sim 0.4\%$)
even for a singly quantized excitation, and grows as the square
root of the number of excitations.} \label{fig5}
\end{figure}

We stress that, in this linear response analysis for $q<q_c$, the
predicted fringe visibility is proportional to $c_q$, which is in
turn proportional to $V_B(u_q+v_q)$. By contrast, at higher
momentum ($q>q_{c}$), the number of observed atoms in the
excitation cloud is linear in the number of excitations, given by
$|c_q|^2$, and quadratic in the quasiparticle amplitudes. This is
consistent with the results of our GPE simulations, where the
fringe visibility at low $q$ and the counting visibility at high
$q$ are found to be proportional to $V_B$ and $V_B^2$,
respectively.

In our experiments and simulations, we typically excite $\sim10^2$
quasiparticles. However, by extrapolating the expected fringe
visibility to lower Bragg intensities, one finds that it remains
sizeable even in the range of single quantized excitations. We
compare (see Fig. 5) the fringe visibility at $qa_{\rho}=0.31$
(circles) with the counting visibility at $qa_{\rho}=2.03$ (boxes)
for our experimental parameters \cite{counting}. The fringe
visibility at low $q$ is predicted to be over two orders of
magnitude more sensitive. This implies a sensitive heterodyne
technique to observe few, and possibly single quasi-particle
excitations.

In conclusion, we discuss and implement a novel fringe visibility
technique to observe the in-situ matter-wave interference of low
momentum excitations with the condensate. We measure the
excitation spectrum with high resolution and analyze the
possibility of singly quantized excitation detection with this
technique.

This work was supported in part by the Israel Ministry of Science,
the Israel Science foundation and the DIP foundation. We thank J.
Steinhauer for his experimental contribution.


\begin{references}
\bibitem{comment1} We note a discrepancy in the axial size of the
expanded condensate as compared to the GPE results, probably due
to experimental imperfections in the release process. The effect
on the radial expansion and on the experimental observables
discussed here is negligible.
\bibitem{offset} In the experimental data (and GP simulation) of Fig. 4 we note an offset due to background signal and a small overlap of the
sidelobes with the central lobe in Fourier space.
\bibitem{note-adiabaticity} In the case of the expanding homogeneous
gas model adiabaticity has been shown to fail for $qa_r < q_{a}a_r
\sim \eta^{-1/2}$ \protect\cite{theory_fringes}, corresponding to
$qa_r<0.3$. However, numerical analysis of the expanding
inhomogeneous system indicates that adiabaticity fails only at an
even lower $q$.
\bibitem{counting} The counting visibility at high $q$ is defined as
the ratio between the number of counted atoms in the excitation
cloud, which is equal to the number of quasiparticles, and  the
total number of atoms $N$.
\end{references}
\end{document}